\documentclass[11pt,tightenlines,eqsecnum,floats,aps,nofootinbib,prd,showpacs]{revtex4}
\usepackage{hyperref}

\usepackage{amsmath,amssymb,amsfonts,amsthm,amscd}
\usepackage{graphicx}
\usepackage{enumerate}
\usepackage{colordvi}
\usepackage{units}
\usepackage{epsfig}
\usepackage{natbib}
\usepackage{enumerate}
\usepackage{colordvi}
\usepackage{multirow}
\usepackage{afterpage}
\usepackage{pdflscape}

\def\be{\begin{equation}}
\def\ee{\end{equation}}
\def\ba{\begin{eqnarray}}
\def\ea{\end{eqnarray}}

\def\f{\frac}

\def\mr{\multirow}
\def\mc{\multicolumn}

\begin{document}

\title{Anisotropic Matter in Cosmology: Locally Rotationally Symmetric Bianchi $I$ and $VII_o$ Models}
\author{David Sloan}
\email{david.sloan@physics.ox.ac.uk}
\affiliation{Beecroft Institute of Particle Astrophysics and Cosmology, Department of Physics,
University of Oxford, Denys Wilkinson Building, 1 Keble Road, Oxford, OX1 3RH, UK}

\begin{abstract}
We examine the behaviour of homogeneous, anisotropic space-times, specifically the locally rotationally symmetric Bianchi types $I$ and $VII_o$ in the presence of anisotropic matter. By
finding an appropriate constant of the motion, and transforming the equations of motion we are able to provide exact solutions in the presence perfect fluids with anisotropic pressures. The 
solution space covers matter consisting of a single perfect fluid which satisfies the weak energy condition and is rich enough to contain solutions which exhibit behaviour which is qualitatively 
distinct from the isotropic sector. Thus we find that there is more `matter that matters' close to a homogeneous singularity than the usual stiff fluid. Example metrics are given for cosmologies 
whose matter sources are magnetic fields, relativistic particles, cosmic strings and domain walls.  \end{abstract}

\pacs{04.20.Dw,04.60.Kz,04.60Pp,98.80Qc,04.20Fy}
\maketitle

\section{Introduction}

We will examine the geometrical behaviour of the type $I$ and $VII_o$ Bianchi models subject to matter which exerts anisotropic pressures. It is usual in cosmology to treat 
pressure as an isotropic function of energy density. There exist a number of physically relevant matter sources for which the pressure is not isotropic - obvious examples include
cosmic strings and magnetic fields which have a clear preferred direction at each spatial point. However it is normally assumed that by appealing to a coarse-graining of such fields
one can ignore any directional preference and deal with the isotropized approximation \cite{Hawking:1965cc}. Here we will show that there are interesting anisotropic physical phenomena which are relevant and
qualitatively distinct from their isotropic counterparts. 

Although one may assume that the large-scale dynamics of the universe should depend only on long-wavelength modes over which an isotropic approximation should hold, this
kind of reasoning breaks in the neighborhood of singularities such as the big bang. Here we expect a significant contribution to come from ever-shorter wave modes as energy
densities are ever increasing. Further, approaches to understanding the nature of singularities such as the BKL conjecture \cite{BKL,AR} posit that dynamics becomes dominated entirely by 
local physics \cite{Ellis:2002hn}, not global phenomena, and therefore any appeal to coarse-grained isotropization is unwarranted. 

We consider locally rotational symmetric (LRS) Bianchi models of cosmology. Under this assumption we can reduce the number of variables that describe a metric considerably.
The LRS condition is of course a simplification of the fully anisotropic sector in which the Bianchi models sit.\cite{Ellis} However this simplification still results in a space of solutions rich 
enough to capture singularities not encountered in the isotropic sector, and is justified when dealing with matter such as radiation whose pressure is axially symmetric, and thus
the effects on geometry can be aptly described by identifying a rotational symmetry about the preferred axis. There are in fact a number of physically interesting scenarios
which are described by matter sources with a preferred axis. Topological defects such as domain walls and cosmic strings, which can arise as a result of spontaneous symmetry
breaking, have a strong anisotropic effect on the geometry. It is usual to treat such model such defects as a gas with no overall directional dependence, however it is clearly not
valid to continue this approximation in a close neighborhood of the defect itself.  

An oft quoted mantra of the oscillatory approach to singularities is that `matter doesn't matter' \cite{Uggla:2003fp,AHS} with the specific exception of a stiff fluid (or massless scalar field). In the isotropic matter case it
is simple to see that in the effect Friedmann equation the anisotropic shear grows as the sixth power of the scale factor, and hence it is apparent that this will dominate over any matter
source whose effect grows more sedately. However, this is not the case when one considers anisotropic matter - in such cases the anisotropic pressures can stop a collapse in one 
directions whilst continuing to allow it in others as we will see in section \ref{Models}, there are physically interesting matter sources which lead to types of singularity which are 
completely distinct from the vacuum cases \cite{Hawking}. Thus we will observe that there are indeed a variety of types of `matter that matters'. 

LRS Bianchi systems have been previously examined - Calogero and Heinzle provided a description of the dynamics of both type $I$ space-times \cite{Calogero:2007wc,Calogero:2007qz} and, in a tour-de-force paper \cite{Calogero:2009mi}
examined all spatially homogeneous LRS space-times with anisotropic matter. These efforts centred around the dynamical systems approach to cosmology and describe the
asymptotic behaviour of systems in terms of a `dominant variable' which is a combination of the Hubble parameter and anisotropic curvatures. Here we will present  
\emph{exact} solutions for single anisotropic fluids which describe the behaviour of one scale factor as a function of the other\footnote{Note the LRS condition means 
that there are only two independent scale factors, the third being identical to the second.}. The precise behaviour of these exact solutions allows us a more detailed view of 
the approach to singularities and the relevant physical scalings of quantities such as curvature and energy density. Likewise we will show that there exists solutions for which
the volume becomes zero at finite and even zero energy density, and yet the dynamics is not vacuum dominated. 

The paper is laid out as follows: In the next section \ref{Setup} we will introduce the physical setup which for the system under examination. Then in section \ref{Transform} we
will show the transformations necessary to obtain a constant of motion and use this constant to find a complete set of solutions. In section \ref{Physics} we examine the solutions
in their physical context, and provide an analysis of their geometrical structures, singularities and the behaviour of energy density, which is followed by some physically 
interesting matter models in \ref{Models}. Finally we present some discussion in section \ref{Discussion}.

\section{Setup}\label{Setup}

We begin with a four dimensional manifold which can be decomposed into the product of a spatial interval and a three-manifold: $M= I \times M_3$. We will be dealing with space-times
that are spatially homogeneous, the Bianchi models, and hence we follow the Bianchi classification, which is based upon the behaviour of Killing vector fields $\xi$.  Thus we can represent all 
such models by the structure constants defining the group of isometries of $M_3$: For structure constants $C^c_{ab}$ defined by
\be [\xi_a , \xi_b] = C^c_{ab} \xi_c \ee
we decompose $C^c_{ab}$ into its components:
\be C^c_{ab} = \epsilon_{abd} n^{cd} + a_a \delta^c_b - a_b \delta^c_a \ee
we can further make the restriction that $a=(a,0,0)$, and describe $n$ in terms of its eigenvalues $n_i$. Space-times for which $a$ is zero are known as the class A models, for which there
exists a well-defined Hamiltonian formalism. We further exploit a symmetry of the systems to identify the eigenvalues of $n$ up to a relative choice of sign, therefore we are left with five separate
models which satisfy the LRS condition, which forces $n_2 = n_3$. The classification is:

\begin{table}[htp]
\begin{center}
\begin{tabular}{| c | c | c |}
\hline
Type & $n_1$ & $n_2$  \\
\hline
 $I$ & $0$ & $0$  \\
 $II$ & $1$ & $0$  \\
 $VII_0$ & $0$ & $ 1$ \\
 $VIII$ & $ -1$ & 1 \\
 $IX$ & $1$ & $1$ \\
 \hline
\end{tabular}
\end{center}
\caption{The LRS Bianchi class A models}
\label{Classification}
\end{table}%

We thus consider metrics which can be decomposed into the form:
\be ds^2 = - dt^2 + q_{ij} (t) W^i W^j \ee 
in which the $W^i$ denote the one forms describing spatial translations on our given three metric $q$. These represent the homogeneous Bianchi models, which we further
 reduce to be diagonal, and satisfy the LRS condition. We are left with components of the metric $q_{11} = a(t)$ which we shall term the `principal' scale factor, and $q_{22} = q_{33} = b(t)$, termed the `secondary' scale 
 factor, with all other components of $q$ set to zero. 

The classification is invariant under reversing of all signs chosen, which simply corresponds to a choice of right- versus left-handed frame. 
Similarly the overall scale of the numerical values of the $n_i$ are unimportant - the system is invariant under a rescaling of $\{n_i,n_j\} \rightarrow \{\eta n_i, \chi n_j\}$ for any
positive real values of $\eta$ and $\chi$. 
We could have equivalently defined the structure constants in \ref{Classification} with the opposite sign and recovered a description of the 
same space-time as a result. We will consider matter in the form of an anisotropic perfect fluid. By fixing the momentum constraint of our system,
 we can choose to align our frame such that this fluid has zero velocity. As such our stress energy tensor is given in terms of direction dependent
  equations of state, relating the anisotropic pressure to energy density:
\be T^i_j = diag(\rho, w_1 \rho, w_2 \rho, w_2 \rho) \ee
We shall assume that the matter content of our models consists of a single perfect fluid, which may have anisotropic pressures. As such
we will fix $w_1$ and $w_2$ to be constants. 

The geometry of our models allows us to calculate the non-zero elements of the Ricci tensor, $R^i_j$, and its trace $R$ as:
\be R^1_1 = \f{n_1^2 a^2}{2b^4} \:\:\:\: R^2_2 =R^3_3 = \f{n_1 n_2}{b^2} - \f{n_1^2 a^2}{2b^4} \:\:\:\: R = \f{2 n_1 n_2}{b^2} - \f{n_1^2 a^2}{2b^4} \ee
The non-zero extrinsic curvatures $k^i_j$ and their trace $k$ are given:
\be k^1_1 = - \f{\dot{a}}{a}  \:\:\:\:  k^2_2 =k^3_3 = -\f{\dot{b}}{b} \:\:\:\: k = -\f{\dot{a}}{a} - 2 \f{\dot{b}}{b} \ee 
And we therefore obtain the equations of motion for the scale factors:
\ba \f{\ddot{a}}{a} &=& -\f{n_1^2 a^2}{2b^4} -2\f{\dot{a} \dot{b}}{a b} + \f{\rho}{2}(1+ w_1 - 2w_2) \\
      \f{\ddot{b}}{b} &=&  \f{n_1^2 a^2}{2b^4}  -\f{\dot{a}\dot{b} }{a b}   -\f{n_1 n_2}{b^2} -\f{\dot{b}^2}{b^2} + \f{\rho}{2}(1-w_1)  \label{master} \nonumber \ea
Note that in the isotropic limit, $a=b$ and $w_1=w_2$ and we would recover the familiar Raychaudhuri equation at this point, written in terms of the matter
content. 

Our system is subject to constraints. Since we are working with homogeneous models, the diffeomorphism constraint is satisfied, and the scalar constraint
results in: 
\be R + K^2 - k_i^j k_j^i = 2 \rho \ee 
In terms of our dynamical variables, this can be expressed as:
\be \rho = 2\f{\dot{a}\dot{b}}{ab} +\f{\dot{b}^2}{b^2} -\f{n_1^2 a^2}{4b^4} +\f{n_1 n_2}{b^2}  \label{RhoG} \ee
In the FLRW case (i.e. the isotropic limit) again we see that this is simply the Friedmann equation relating the square of the Hubble rate to the energy density
and curvature of the system in consideration (note that the choice of sign of the product $n_1 n_2$ is what is important in terms of curvature). 

Combining this with our equations of motion for the scale factors, we come to the complete equations of motion satisfying the constraint: 
\ba \label{MainEoM} \f{\ddot{a}}{a} &=& (w_1 -2w_2 -1) \f{\dot{a}\dot{b} }{a b} + \left( \f{1+w_1 - 2w_2}{2} \right) \left(\f{\dot{b}^2}{b^2} + \f{n_1 n_2}{b^2}\right) + \left(\f{2w_2 -w_1 -5}{8}\right) \f{n_1^2 a^2}{b^4} \nonumber \\
      \f{\ddot{b}}{b} &=& -w_1  \f{\dot{a}\dot{b} }{a b} - \left( \f{1+w_1}{2} \right)  \left(\f{\dot{b}^2}{b^2} + \f{n_1 n_2}{b^2}\right)  + \left(\f{3+w_1}{8}\right)  \f{n_1^2 a^2}{b^4}  \ea
This systems exhbits a symmetry under rescaling which we expect to see from the homogeneity of the background geometry - rescaling both scale factors by arbitrary real numbers 
does not change dynamics, if corrected for by an appropriate re-scaling of $n_1$ and $n_2$. In particular, in the cases where $n_1$ is zero we should expect to see an arbitrary
choice of scale factor - rescaling $a$ and $b$ does not change the equations of motion, as these only appear in ratio with their time derivates. In other cases this factor will be
 fixed only up to a choice of the $n_i$. To make this explicit, consider the transformation 
\be \{a,b,n_1,n_2\} \rightarrow \{\mu a, \nu b, \nu^2 \mu^{-1} n_1, \mu n_2 \} \ee
for arbitrary positive real numbers $\mu$ and $\nu$. Under this transformation, both scale factors change, yet the equations of motion remain invariant (recall that the choice of $n_1$ and $n_2$ was arbitrary, all that matters is their (relative) sign and whether or not these are zero). 

In the isotropic limit, each of these equations reduce to the familiar Raychaudhuri equation for a single perfect fluid which relates the acceleration of the scale factor to 
the pressure and energy density. Let us denote the isotropic versions of quantities by the subscript $I$. 
\be \f{\ddot{a_I}}{a_I} = - \left(\f{3w_I + 1}{2}\right) \left(\f{\dot{a_I}^2}{a_I^2} + \f{N }{8 a_I^2}  \right) \ee
Here $N$ represents the `isotropized' value of the $n_i$ which is the product $n_1 n_2$. Note that this is of course
freely rescaled by a choice of overall scale of the $n_i$ and equivalent to the $k$ in FLRW models - the only important
thing about it is its sign, the value can be chosen by an arbitrary choice of coordinates. It may be unfamiliar to see the 
presence of a curvature term in the Raychaudhuri equation, however this arises as a result of the Friedmann equation and
the relationship between energy density and pressure. 

\section{Methods of Solution}\label{Transform}

Let us consider the Bianchi type $I$ and $VII_0$ space-times. These are spatally flat, and hence each contains the $k=0$ Friedmann-Lem\^aitre-Robertson-Walker 
solutions as their isotropic limit. For each of these, $n_1 = 0$ and hence the equations of motion (\ref{MainEoM}) reduce considerably in their complexity, and become 
tractable upon identifying constants of the motion.  In each case our strategy for finding a solution will be the same - first we use 
a transformation of variables which allows us to find a constant of the motion from the second equation of (\ref{MainEoM}) which will be expressed in terms of the set
$\{a,b,\dot{b}\}$. Then we use this constant to transform derivatives with respect to $t$ into derivatives with respect to $b$. We note that from equation (\ref{RhoG})
it is apparent that in the presence of matter, there cannot be a turning at of the secondary scale factor at which it takes a finite value. Since $b$ is monotonically non-decreasing 
and lies in the interval $[0,\infty)$ this will parametrise the complete space-time. Finally we solve the resulting equation of motion to find $a(b)$, and examine the
singular behaviour thereof. We will generally consider space-times in which $b$ is increasing - it is a trivial transformation to reverse the time direction and consider 
space-times in which $b$ is always decreasing. On first inspection, this transformation may seem counterintuitive: It is usual when attempting to solve complex 
coupled equations of motion to solve parametrically - providing $\{x(\eta),y(\eta)\}$ in place of $x(y)$. However in this case we shall see that the extraneous structure
which is used in describing the geometry, namely the time variable, is actually an impediment to finding a solution. Once a full parametrisation $x(y)$ or $y(x)$ is given
we can of course return to introduce an external parameter which can be thought of as time, and we shall do this when presenting particular physical solutions in section \ref{Models}. 

\subsection{Finding a constant}

In order to find a constant of the motion, we will perform a change of variables that makes the equations more malleable. For completeness, here we will show the
 general case of the transformation made, then specialise this to the specific context of the LRS systems under consideration. First note that one of the equations of 
 motion is of the form 
\be \f{\ddot{y}}{y} = \alpha \f{\dot{y}\dot{x}}{yx} + H(x,y,\dot{y}) \ee
Our goal is to reduce the order of the differential equations, and hence we make a substitution designed to eliminate the second derivative term. One can define
 the variable $m=x \dot{y}^\beta$. In doing so we find that 
\be \dot{m} = \dot{x}\dot{y}^\beta + \beta x \dot{y}^{\beta-1} \ddot{y} \ee
and hence we can rearrange the equation to recover the product term in our equation of motion. Thus we find
\be \f{\dot{x}\dot{y}}{xy} = \f{\dot{m}\dot{y}}{my} - \beta\f{\ddot{y}}{y} \ee
We can then choose $\beta=-1/\alpha$ and achieve the goal of reducing our differential equation by removing the second derivative term in $y$, resulting in:
\be \alpha \f{\dot{m}\dot{y}}{my} + H(m \dot{y}^{\f{1}{\alpha}} , y,\dot{y}) = 0 \ee
If the function, $H$ can be separated such that $H(m \dot{y}^{-\beta} ,y, \dot{y}) = M(m)Y(y,\dot{y})$ we can integrate to leave the constant of integration and a
 constant of motion for the system. In the systems we are considering here, the function $M=1$ and hence the separation is trivial. On doing so we find our constant $J$ is given by
\be J = \dot{y}^{1/\alpha} x \exp[\int \f{y Y(y,\dot{y})}{\alpha \dot{y}} dt] \ee
We have established now the general constant of motion that will arise in such systems. Let us now specialise to those systems described in (\ref{MainEoM}). Here the equation for Y takes the form (for some constant $\lambda$ composed of the coefficients from \ref{MainEoM})
\be Y(y,\dot{y}) = \lambda \f{\dot{y}^2}{y^2} \ee
and so our integral is performed easily to yield a constant, $L$ given in terms of the set $\{x,y,\dot{y}\}$. Since we will be using this constant to eliminate derivates with 
respect to time later, we can raise the constant to a suitable power to leave a constant which is linear in $\dot{y}$;
\be J = L^{-\alpha}=x^{-\alpha} y^{-\lambda} \dot{y} \label{Const} \ee
Note that our analysis would appear break down if $\alpha=0$. This is indeed the case for a subset of the space-times described by (\ref{MainEoM}), and hence we will 
have to treat these solutions separately. Fortunately, however, such systems are easily directly integrable, since the equation of motion for $x$ becomes independent of $y$.
In such cases we can solve directly the equation of motion, resulting in
\be y=y_o ((1-\lambda)(t-t_o))^{\f{1}{1-\lambda}} \label{alpha0} \ee
From this point we could continue by transforming the equation of motion for $x$ to remove the $t$ dependence in favour of dependence on $y$. However, from our
solution it is simple to note that we can again recover a constant which is consistent with $J$ extended to the $\alpha=0$ case; $J=\dot{y}y^{-\lambda}$. Hence in 
either case we can continue our analysis using this constant of motion to simplify the equation of motion for $x$. 

We should note further at this point that the existence of our constant $J$ given by equation (\ref{Const}) means that $y$ must be monotonic throughout - the existence
of the constant is incompatible with the vanishing of $\dot{y}$ except where the other parameters become singular. Since these parameters $\alpha$ and $\lambda$ are
simply constants themselves, this can only occur at points where $x$ or $y$ vanish or become infinite (depending on the signs of $\alpha$ and $\lambda$). We are therefore
justified in making a transformation of the equations of motion throughout their entire range. 

\subsection{Transforming the equation}

Now that we have access to a constant of the motion, we are in position to both eliminate terms involving derivatives of $b$ from the equation of motion for $a$, yielding a
more malleable equation to solve. The equation of motion for $a$ is of the form:
\be \f{\ddot{x}}{x} = \mu \f{\dot{x}\dot{y}}{xy} + \nu \f{\dot{y}^2}{y^2} \ee
Making our substitution we find this results in
\be \f{\ddot{x}}{x} = \mu \f{J \dot{x}}{x^{-\alpha+1}y^{-\lambda+1}} + \nu \f{J^2}{x^{-2 \alpha}y^{-2\lambda+2}} \ee
The powers of $J$ that are involved is $2-n$ where $n$ is the number of time derivatives acting upon the $x$ terms which the $J$s multiply. It is therefore possible to make
a change of variables; using the fact that $J dt = x^\alpha y^\lambda dy$ we can eliminate time derivatives in favor of those in terms of $y$ (henceforth denoted by a prime). Upon doing so, and after some algebraic
manipulation to simplify matters we are left with
\be\f{x''}{x} y^2  = -\alpha  \f{x'^2}{x^2}y^2 +(\mu-\lambda)\f{x'}{x}y + \nu \ee
Which has general solution:
\be \label{Genx} x=x_o y^{-\frac{\sqrt{4 (\alpha +1) \nu +(-\lambda +\mu +1)^2}+\lambda -\mu -1}{2 (\alpha +1)}} \left(C+y^{\sqrt{4 (\alpha +1) \nu +(-\lambda +\mu +1)^2}}\right)^{\frac{1}{\alpha +1}}\ee
for free choices of $x_o$ and $C$, whose ranges will be determined by physical considerations (such as positivity of scale factor and energy). It is also apparent that special attention needs to be 
paid when $F=4 (\alpha +1) \nu +(-\lambda +\mu +1)^2=0$ or $\alpha=-1$. In the former case, the space of solutions spanned by \ref{Genx} is reduced to a single dimension, with solution 
$x_1=x_o (1+\alpha)y^{\frac{-\lambda +\mu +1}{2 (\alpha +1)}} $ and thus we should look for a second solution. By making the substitution $x_2 = l x_1$ we find that $l(x)$ satisfies
\be \frac{l''}{l} y^2+\alpha \frac{l'^2}{l^2}y^2+\frac{l'}{l}y = 0 \ee
and hence we can easily recover a general solution in this case, 
\be x=x_o (\alpha +1) (C+(\alpha +1) \log (y))^{\frac{1}{\alpha +1}} y^{\frac{-\lambda +\mu +1}{2 (\alpha +1)}} \label{Logx} \ee
again we note that there is a special sub-case in which $\alpha=-1$. Here we find the solution takes the form $x=x_o y^C$. 

We can now turn our attention to the second special set of solutions - those for which $\alpha=-1$. In this case we can solve our equation of motion to obtain
\be x=x_o e^{Cy^{-\lambda +\mu +1}} y^{\frac{\nu }{\lambda -\mu -1}} \label{a1} \ee

Thus we have a complete set of solutions for systems such as those described in \ref{MainEoM}. From solutions we have obtained it is apparent
that our space-times will display qualitatively distinct behaviours in some cases which will be determined by the free parameters describing the 
models - in our case these will be combinations of the anisotropic pressures $w_1$ and $w_2$. The space of solutions is unsurprisingly richer 
than that of the purely isotropic sub-systems. In the following section we will discuss the geometries of solutions in the richer space and analyse
their singularities.

\section{Physical Solutions}\label{Physics}

We will now take the solutions obtained in the previous section and examine their nature in the physical situation described by equation (\ref{MainEoM}). The
complete set of solutions to these systems is summarized in table (\ref{Total}) which gives the exact behaviour of the solutions together with their singular
points and the behaviour of energy densities. 

The space of solutions to these models depend on the values of $w_1$ and $w_2$, the anisotropic pressures. From inspection of the equations of 
motion we can identify the parameters in equation \ref{Genx}. In particular we note that 
\be \alpha = -w_1 \quad \lambda=-\f{1+w_1}{2} \quad \mu= w_1-2w_2-1 \quad \nu= \f{1+w_1-2w_2}{2} \ee
in particular, we note that the parameter which is particularly important for distinguishing solutions will be determined by
\be F=\sqrt{4(\alpha+1)\nu+(1-\lambda+\mu)^2} = \f{3}{2}+\f{w_1}{2}-2w_2 \ee
 We will use this combination to split the space of solutions into the cases where $F > 0$, $F=0$ and $F<0$. Note that in the isotropic cases we would always have $F \geq 0$ so long as $P \leq \rho$. This is easily extended
in the anisotropic case by setting $w_1 =0, w_2=1$ for example, and hence the pressure is always less than or equal to energy density and yet $F=-1/2$. In fact $F$ runs from -1 to 4 in the anisotropic case, going beyond the 
range for which the isotropic equivalent, $\f{3}{2}(1-w_I)$, which would run from 0 to 3. Thus we should initially expect that any qualitative behaviour distinct from that exhibited by FLRW cosmologies should occur for $F$ outside
of this range. 

The constant of motion, $J$ given by equation (\ref{Const}) can be expressed in terms of the scale factors and the anisotropic pressures. By substituting the values of $\lambda$ and $\alpha$ we see that the constant is now
given:
\be J = a^{w_1}b^{\f{w_1+1}{2}}\dot{b} \ee
which is independent of the value of $w_2$. In the isotropic sector this reduces to the Friedmann equation in the presence of a single perfect fluid; setting $a=b$, we see that the $J$ becomes
\be J=\dot{a}a^{\f{1+3w}{2}} \rightarrow \f{\dot{a}^2}{a^2} = \f{J^2}{a^{3(1+w)}} \ee
as we would expect for a single fluid with equation of state $P=w\rho$ in FLRW cosmologies. 

For each set of solutions we will give the complete behaviour and describe the singularities encountered. Of particular interest will be the energy density for each of these solutions, which is given:
\be \rho = \f{1}{a^{2w_1}b^{w_1+2}} \left(2 \f{a'}{a} + \f{1}{b} \right) \label{Rho} \ee
Note that energy density should always be positive. This will determine the range of values of constants in our solutions which give rise to physical solutions. In the systems we consider the matter content is always described
by a single perfect fluid with a fixed equation of state. Therefore pressure can only diverge when energy density diverges, and so can be found trivially once the energy density is calculated. Typically we will fix $a_o$ and $b_o$
to unity in order to simplify matters, and will therefore treat $\rho$ up to a constant that depends on these arbitrary choices.

This section will be split into subsections dealing first with the case $w_1=1$, which is a special set of solutions then the general solutions depending on the sign of $F$. This is then followed by table \ref{Total} which summarises
all cases. 

\subsection{Solutions for which $w_1=1$}

As was noted in section \ref{Transform}, the case in which $\alpha=0$ (equivalently $w_1=0$) has a distinct space of solutions from the rest. First
let us note that $F=2-2w_2$, and so our space of solutions separates depending on whether $w_2$ is greater than, less than or equal to 1. 
\footnote{Our energy conditions do not allow for $w_2 >1$, however we present the result simply for completeness.} Let us first consider $F=0$. Using
equation \ref{a1} we can see that the behaviour of the principal scale factor, $a$ can be given as
\be a = a_0 b^A \label{a10}  \ee
(note that $A$ is a constant of integration). We can explicitly evaluate $\rho$ to see behaviours at singularities and ranges of validity of solutions. The case $F=0$ contains the only isotropic subsector in this case. This should be unsurprising as the pressures are distinct, therefore the 
presence of any matter of this type will break isotropy. Here we find 
\be \rho = \f{1+2A}{b^{2A+4}} \label{Rho10} \ee
for arbitrary choice of $A>-1/2$ (so that energy remains positive throughout). We see that the energy density scales as the inverse square of the volume, as expected in the case of a stiff fluid. Note that in the case where $A=-1/2$, the vacuum case, we indeed 
recover the Kasner solution, $a=b^{-1/2}$, i.e.
\be ds^2 = - dt^2 +t^{-2/3} dx^2 + t^{4/3} (dy^2 +dz^2)  \label{LRSKasner} \ee
Since $F=0$ implies that $w_2=1$ this situation corresponds to that of a stiff fluid, or scalar field. This solution is of particular interest
as it is central to the terms that `matter' in the BKL conjecture, as in the case of isotropic matter pressures this is the only term which
contributes at the same order of volume as the anisotropic shears in the equation for the isotropized Hubble parameter. By making the
ansatz that $b = t^\gamma$ we find that $\gamma=(A+2)^{-1}$, we can describe the full space of solutions by
\be ds^2 = -dt^2  + t^{2-4 \gamma} dx^2 +t^{2 \gamma} (dy^2 +dz^2) \ee
in which $\gamma$ runs from $0$ at $A \rightarrow \infty$ to $-2/3$ at $A=-1/2$. Here we recover the Bianchi I condition that the sum
of the exponents of the time variable is equal to unity. From this we can trivially see that there are distinct
spaces of solutions determined by the $A$ - if $A>0$  the principal scale factor is zero when the secondary vanishes, and both go to
infinity together. However, if $A=0$ we see that the the principal scale factor stays constant whilst the secondary expands to infinity. Finally
for $A<0$  we see a system that begins with a `cigar' type singularity in which the principal scale factor is infinite and the secondary zero, to
a `pancake' in which the secondary scale factor goes to infinity whilst the primary tends to zero. In all these cases the initial singularity has 
infinite energy density at $b=0$ and zero final energy density at $b \rightarrow \infty$.  Here we find that, in parallel to the BKL type 
analysis, the nature of the initial time singularity is dependent on both geometrical and matter degrees of freedom - `matter does matter'.

Let us now turn our attention to the cases in which $F \neq 0$. As we shall see in the next section, this contains the ultra-relativistic homogeneous particle
steam. Here we find the principal scale factor is given by
\be a= a_o \f{\exp(Ab^F)}{\sqrt{b}} \label{a1p} \ee
It is readily apparent that for such solutions there cannot exist an isotropic sector - this is unsurprising as the conditions on matter for the the anisotropic 
pressures to differ from one another. For these solutions the energy density is given by an exponential function of the secondary scale factor $b$, raised to
the appropriate power ($F$), and hence we are guaranteed that the energy density will vanish at large values of $b$. 
\be \rho =  2AF \exp(-2Ab^F)b^{F-3} \label{Rho1p} \ee
Since we require that the energy density is positive throughout, both $F$ and $A$ must have the same sign. Again we
recover the LRS Kasner solution (\ref{LRSKasner}) in the vacuum case. It is therefore apparent that in the limit as the 
secondary scale factor tends to zero we see that the principal scale factor becomes infinite, as the numerator tends to 
a finite (positive) value. In the large $b$ limit, we see that the primary scale factor tends to infinity. Between these two 
limits there exists a minimum value of the primary scale factor at:
\be b=(2AF)^{-\f{1}{F}} \quad a=(2AFe)^\f{1}{2F} \label{b1p} \ee
wherein $e$ is the base of the natural logarithm.

From equation (\ref{Rho1p}) the asymptotic behaviour of the energy density is easily ascertained. At large $b$, $\rho \rightarrow 0$ for all values of $F$. 
At low $b$, $\rho \rightarrow \infty$ if $F<3$, $\rho \rightarrow 0$ if $F>3$ and if $F=0$ $\rho$ becomes a constant determined by $A$. 
In the case where $F>3$ the energy density both starts and ends at zero. The maximum can readily be found, and occurs when 
$b^F = \f{F-3}{2AF}$, and is the sole turning point of energy density. Thus we find that in these cases both the extreme behaviours have
asymptotically zero energy density - the initial singularity is still vacuum dominated. 

Here we again see behaviour that is qualitatively distinct from the isotropic sector: If $w_2<-1/2$ (i.e. $F>3$) we find that the
initial time singularity happens with vanishing energy density. This scenario highlights the inadequacy of the coarse-graining
arguments - had one considered the isotropic coarse-grained equivalent, the isotropic pressure ($w=\f{w_1 + 2 w_2}{3}$)would be 
less than zero and the initial singularity would occur at zero volume with infinite energy density. 

\subsection{Solutions for which $w_1 \neq 1$}

There are many situations in which $w_1$ is not equal to one. First let us consider the general case in which $F$ is
non-zero. Here we know that solutions are given by equation (\ref{Genx}). The space of solutions actually splits into two, based
on the sign of $F$. Treating first the case where $F>0$, we find that for positive $F$, inserting our the relevant parameters
from equations (\ref{MainEoM}) into the system:

\be a=a_o \f{(A+b^F)^{\f{1}{1-w_1}}}{\sqrt{b}} \label{anp} \ee
for a free choice of $A$ and $a_o$. Since we require that the scale factor be positive, if $F>0$ we se that at large $b$ the term in 
containing $b^F$ will dominate, and hence we can keep $a_o$ positive, regardless of the choice of $A$. 

The behaviour of these systems close to the initial singularity depends upon the choice of $A$. In cases where $A$ is positive or zero, the numerator
is everywhere positive, and hence at low values of the secondary scale factor, the principal scale factor will tend to infinity. Hence we observe
a `cigar' type singularity at this point. When $A<0$ we find that at the point where $b^F = -A$ the principal scale factor vanishes, and hence we are
left with a singularity at which one of the scale factors is zero where the other takes a finite value - a `barrel' singularity. 

The principal scale factor has a (unique) turning point if $A<0$ and $1+w_1<2w_2$. In this case $a$ begins at zero at the barrel singularity and expands
to a maximum at
\be b^F = -\f{A(1-w_1)}{1-2F-w_1} \ee
and after this point re-contracts down to zero eventually. Likewise if $A>0$ and $1+w_1>2w_2$ we find that $a$ begins at infinity, reduces down to a minimum
value at the point above, and re-expands to infinity thereafter. We thus note that the asymptotic behaviour of the principal scale factor is determined by $1+w_1-2w_2$ - 
if this is negative we have endless expansion, if positive we recollapse to a pancake singularity, and if this is zero $a$ asymptotes to a finite value as $b$ tends 
to infinity. 
 
We can find the energy density in from equation (\ref{Rho}). On doing so we see that
\be \rho = \f{2 F \left(A+b^F\right)^{-(\f{1+w_1}{1-w_1})}b^{F-3}}{1-w_1} \label{Rhonp} \ee
For large $b$ the asymptotic behaviour is
\be \rho \rightarrow b^{-\f{w_1^2 - 4w_1 w_2 + 3}{1-w_1}} \ee
This always asymptotes to zero for matter obeying the energy conditions (to see this note that the exponent is minimal when $w_2=1$, and in this circumstance
we see that the energy density follows $b^{w_1-3}$, and so always tends to zero). The one notable exception is the when $w_1=-1$ that $\rho=Fb^{3-F}$. The maximal value of $F$ is 3 in this case, which
corresponds to all pressures being the negative of energy density - a cosmological constant. Unsurprisingly in this case we find that the energy density does not change
over time. Hence the energy density at the initial singularity is a constant in this case.

The behaviour of energy density at the initial singularity is somewhat more complex. When $w_1=-1$, with the exception of the case of the cosmological constant \cite{Goliath:1998na}, $\rho$ is always divergent. 
However, when $w_1>-1$ there exist a number of circumstances under which $F>3$ and hence there could be zero energy density at the initial singularity. The behaviour here depends crucially
on the type of singularity - if the singularity is barrel like (i.e. $a$ vanishes at a finite value of $b$, characterised by $A<0$) we see that the energy density is always divergent at this point, following
\be \rho \rightarrow (A+b^F)^{-\f{1+w_1}{1-w_1}} \ee
in which the exponent is always negative. In the case where $A>0$, the behaviour is determined by 
\be \rho \rightarrow b^{F-3} \ee
and hence may tend to a constant when $F=3$ or even zero for $F>3$ (recall that $F$ is bounded above by 4 in the anisotropic case, as opposed to 3 in the isotropic sector). 
Hence 

The case $w=-1$ also contains an isotropic sector when $F>0$: Isotropic matter pressures are achieved when $w_1=w_2=w$, and so $F=\f{3}{2}(1-w)$. Together with our equation
of motion \ref{anp} we see that in such a case 
\be a=a_o \f{ (A+b^{\f{3}{2}(1-w)})^{\f{1}{1-w}}}{\sqrt{b}} \ee
and geometrical isotropy further requires that $A=0$ (and hence $a=b$). The energy density in such a situation is given by 
\be \rho = \f{3(A+b^{\f{3}{2}(1-w)})^{-\f{1+w}{1-w}}}{b^{\f{3}{2}(1+w)}} \ee
again in the geometrically isotropic situation, $A=0$ and so this reduces to the familiar $b^{-3(1+w)}$. 

The geometry of solutions for which $F=0$ is determined by equation (\ref{Logx}). Substituting the factors in terms of their expressions in terms of
anisotropic pressures we find that the principal scale factor is given by
\be \label{an0} a=a_o \f{(A+(1-w_1)\log(b))^{\f{1}{1-w_1}}}{\sqrt{b}} \ee
As $b \rightarrow 0$ we note that $\log(b) \rightarrow  -\infty$. Therefore in these geometries, for any real value of $A$ there will always be a finite value of the 
secondary scale factor at which the primary scale factor vanishes. For large $b$ the denominator will always dominate, and thus the geometry asymptotes to a pancake. 
Between these points there exists a unique turning point at
\be b=e^{\f{2-A}{1-w_1}} \quad  a=a_o 2^{\f{1}{1-w_1}} e^{\f{2(1-w_1)}{2-A}}   \ee 
Thus we find that this is a situation in which the geometry always transitions between two pancake solutions, the first at a finite, non-zero value of the secondary scale factor,
and the second as the secondary scale factor tends to infinity. 

The energy density for such geometries is given:
\be \label{Rhon0} \f{2(A+(1-w_1)\log(b))^{-\f{1+w_1}{1-w_1}}}{b^{3}}  \ee
and we can see that such systems must begin with infinite energy density, as this includes a term that is a negative power of the principal scale factor. The behaviour of $\rho$ 
is dominated at large values of the secondary scale factor by the denominator, as this grows faster than the logarithmic term in the numerator. Hence the energy density
vanishes at the final pancake surface. 

Finally we come to the case in which $F<0$. As noted above, such for these solutions the weak energy condition ensures that the matter cannot be isotropic, therefore we are
 in a regime which qualitatively differs from the isotropic sector. The solutions to the equations of motion are again given by equation (\ref{Genx}). However, in these cases we find
 that the expansion of the square root terms has the opposite sign to that described in the $F>0$ case, and thus we can express our space of solutions as those given by: 
\be a=a_o \f{(1+Ab^F)^{\f{1}{1-w_1}}}{\sqrt{b}} \label{anm} \ee
The energy density of these solutions is given:
\be \label{Rhonm} \rho = \f{2AF(1+Ab^F)^{-\f{1+w_1}{1-w_1}}}{(1-w_1)b^{3-F}} \ee
and thus we observe that the positivity of energy density requires that $A$ be negative. We can therefore immediately infer the behaviour of these solutions in the vicinity of their 
initial singularity. When $b$ is sufficiently small the term in the numerator of equation (\ref{anm}) vanishes, and thus we see that we have a finite pancake initial singularity - the 
principal scale factor vanishes at a non-zero value of the secondary scale factor. Likewise as $b$ becomes large, $a \sim a_o / \sqrt{b}$ and thus the final singularity is an infinite
pancake. Solutions have a turning point at 
\be b^F = -\f{(1-w_1)}{A(1-2F-w_1)} \ee

The behaviour of the energy density is also entirely determined - at the initial pancake singularity, the energy density is infinite, and as the secondary scale factor becomes large,
we see that $\rho \sim 1/b^{3-F}$, and hence tends to zero. Therefore the geometry of these solutions is unusual - the systems starts and ends at pancake singularities, the former
finite, the latter infinite, having a finite extent in the principal direction in between. The energy density of these solutions decreases throughout.

\afterpage{
    \clearpage
    \begin{landscape}
\begin{table}[htp]
\begin{center}
\begin{tabular}{| c | c | c | c | c | c | c c | c | c  c |}
\hline

\mc{2}{|c|}{Sector}		& Scale Factor 							& Energy Density 															& Range 				& Zero Size 			& \mc{2}{|c|}{Infinite Size}  			& Turning Point						& \mc{2}{|c|}{Singular Density}\\
\cline{1-2}
$w_1$		& $F$ 		& $a(b)$ 							& $ \rho(b) $ 															& (A) 				& $ a(0) $ 			& \mc{2}{|c|}{$ a(\infty)$}  			& $a'(b) = 0$ 						& \mc{2}{|c|}{$\rho(0)$}\\
\hline
\mr{3}{*} {1}	& \mr{3}{*}{+}	&\mr{3}{*} {$\f{\exp(Ab^F)}{\sqrt{b}}$}	& \mr{3}{*}{$\f{2AF}{\exp(2Ab^F)b^{3-F}}$}									& \mr{3}{*}{$(0,\infty)$}	&\mr{3}{*}{$\infty$}	& \mc{2}{|c|}{\mr{3}{*} {$\infty$}}	& $b=(2AF)^{-\f{1}{F}}$				& 0 		& $F>3$	\\
			&  			&								&																  	&		      	 		&				&			&				&  $a=(2AFe)^\f{1}{2F}$	& 2AF 	& $F=3$	\\
			& 			&								&																	&		 			&				&			&				&								& $\infty$ 	& $F<3$ 	\\
\hline
\mr{3}{*} {1}	&  \mr{3}{*} {0}	&\mr{3}{*} {$b^A$} 					& \mr{3}{*} { $\f{1+2A}{b^{2A+4}}$}											& $(-\f{1}{2},0)$			& $\infty$ 			& \mc{2}{|c|}{0}					& N/A							&\mc{2}{|c|}{\mr{3}{*}{$\infty$}}	\\
			&			&								&																	& $0$				& Const 			& \mc{2}{|c|}{Const} 				& Always							& 		&		\\
			&			&								&																	& $(0,\infty)$			& 0				& \mc{2}{|c|}{$\infty$}			& N/A							&		&		\\
\hline
 \mr{2}{*} {1}	& \mr{2}{*}{-}	&\mr{2}{*} {$\f{\exp(Ab^F)}{\sqrt{b}}$}	& \mr{2}{*}{$\f{2AF}{\exp(2Ab^F)b^{3-F}}$}									& \mr{2}{*}{$(-\infty,0)$}	&\mr{2}{*}{$0$}	& \mc{2}{|c|}{\mr{2}{*} {$0$}}			& $b=(2AF)^{\f{-1}{F}}$				& \mc{2}{|c|}{\mr{2}{*}{$\infty$}}	\\
			&  			&								&																  	&				     	&				&			&				& $a=(2AFe)^\f{1}{2F}$	&		&		\\
\hline
\mr{3}{*}{$\neq 1$}&  \mr{3}{*} {+}&\mr{3}{*} {$\f{(A+b^F)^{\f{1}{1-w_1}}}{\sqrt{b}}$} 	& \mr{3}{*} { $ \f{2F(A+b^F)^{-\left(\f{1+w_1}{1-w_1}\right)}}{(1-w_1)b^{3-F}}$}	& $(-\infty,0)$			& $b=-A^{\f{1}{F}}$ 	& $\infty$	& $1+w_1>2w_2$		& \mr{3}{*}{$b^F = -\f{A(1-w_1)}{1-w_1-2F}$}& $\infty *$ & $w_1 = -1$	\\
			&			&								&																	& $0$				& $\infty$			& 1		& $1+w_1=2w_2$		& 								&$\infty$& 	$w_1 \neq -1 \:\: A \leq 0$	\\
			&			&								&																	& $(0,\infty)$			& $\infty$			& 0		& $1+w_1<2w_2$		& 								&$ \sim b^{F-3 }$& $w_1 \neq -1 \:\: A > 0	$	\\
\hline
 \mr{2}{*} {$\neq 1$}	& \mr{2}{*}{0}	&\mr{2}{*} {$\f{(A+(1-w_1)\log(b))^{\f{1}{1-w_1}}}{\sqrt{b}}$}	& \mr{2}{*}{$\f{2(A+(1-w_1)\log(b))^{-\f{1+w_1}{1-w_1}}}{b^{3}} $}& \mr{2}{*}{$(-\infty,\infty)$}&\mr{2}{*}{$b=e^{-\f{A}{1-w_1}}$}& \mc{2}{|c|}{\mr{2}{*} {$0$}}& $b=e^{\f{2-A}{1-w_1}}$			& \mc{2}{|c|}{\mr{2}{*}{$\infty$}}	\\
			&  			&								&																  	&		     			 &				&			&				& $a=\f{2^{\f{1}{1-w_1}}}{\sqrt{b}}$	&		&		\\
\hline
\mr{2}{*}{$\neq 1$}&\mr{2}{*}{ -}	& \mr{2}{*}{$\f{(1+Ab^F)^{\f{1}{1-w_1}}}{\sqrt{b}}$}	& \mr{2}{*}{$\f{2AF(1+Ab^F)^{-\f{1+w_1}{1-w_1}}}{(1-w_1)b^{3-F}}$} 			&\mr{2}{*}{$(-\infty,0) $}	&\mr{2}{*}{$b=-A^{-\f{1}{F}}$}	&\mc{2}{|c|}{\mr{2}{*} {$0$}}				& \mr{2}{*}{$b^F = -\f{1-w_1}{A(1-w_1-2F)}$}	& $\infty$ & $w_1 \neq -1$ \\
			&			&										&															&					&				&			&				&					&$F A^{\f{F-3}{F}}$&$w_1 = -1$ \\		
\hline	
\end{tabular}
\end{center}
\caption{All the type $I$ and $VII_o$ with ranges and behaviours. Note that $a(0)$ is used to describe the value of $a$ at $b=0$ or the value of $b$ for which $a=0$. Likewise $\rho(0)$ describes the value of $\rho$ at either $a=0$ or $b=0$. 
The behaviour of the scale factors is determined up to a free choice of $a_0$ and $b_o$ - a coordinate choice. Similarly the energy density is only determined up to this free choice in the model, which amounts to a choice of units. ${}^*$ Note that the energy density in the case $F>0$ and $w_1=-1$ is constant when $F=3$ (i.e. $w_2=-1$, the case of the cosmological constant).}
\label{Total}
\end{table}
\pagebreak
\end{landscape}
    \clearpage
    }

\section{Matter Models}\label{Models}

There are a wide number of matter models which are anisotropic at the fundamental level, yet have been treated as isotropic gasses when considering cosmological effects. 
This treatment is justified in the large scale approximation, in which short range fluctuations are thought to average out in some manner. However, on considering the 
approach to a singularity, it is clear that short range physics is important. In the context of local physics dominating dynamics, (such as that described by quiescent cosmology)
this approximation becomes invalid. It is therefore necessary to consider anisotropic solutions. Some analysis of these models is presented in \cite{Aluri:2012re} which considers
the impact of such matter sources on cosmological observations. 

\subsection{Cosmic Strings}

A clear example of this is the case of topological defects, such as cosmic strings and domain walls \cite{Vilenkin:1984ib,Hindmarsh:1994re}. In each of these cases there is a preferred axis (along the string or 
perpendicular to the wall) about which we should expect our geometry to behave isotropically, but along which we should see different motion. It is in these contexts that 
the LRS ansatz is justified. The local dynamics near a cosmological string is determined by the energy density of the string itself, which we will assume is describe in a
geometry in which the string is at rest \cite{Allen:1990tv}. For such a system the anisotropic pressure
are such that $w_1=-1$ and $w_2=0$, and hence $F=1$. These solutions fall into the category described in equation(\ref{anp}) - the principal scale factor behaves as 
$a = a_o \sqrt{\f{A}{b} +1}$, and there is a more complex set of singularities depending on the value of $A$ - at late time $a$ approaches a constant value, and at early
times $a$ diverges to infinity in a cigar-like singularity. This is significantly different from the isotropic case, where the equivalent behaviour is described by a string gas,
in which the averaged pressure gives $\overline{w}=-1/3$. Similarly we see that the energy density, given by equation(\ref{Rhonp}) follows $b^{-2}$. At first this may seem surprising - 
in the isotropic case we expect the string energy to fall off with average scale factor to the fourth power. However, by stretching our space-time in the principal direction, we
should increase the string energy in the region such that the density remains constant, and therefore the only scaling should be due to a dilution following $b^2$. 
From our constant of motion, it is possible to determine exactly the metric given by these solutions in some circumstances. Since the constant
is $\dot{b}/a$, we can easily invert: 
\be \dot{b} = C a = C \sqrt{1-\f{A}{b}} \ee
and hence we can solve, (setting C = 1 through an appropriate rescaling of the temporal coordinate);
\be t=\sqrt{b(A+b)} - A \log(\sqrt{b(A+b)}) \ee
in the simple case, where $A=0$, this can be inverted, yielding the metric:
\be ds^2 = -dt^2 + dx^2 + t^2(dy^2+dz^2) \ee
From this we explicitly see the singularity as having a `barrel' type - the $y$ and $z$ directions collapse whilst the $x$ direction remains finite. Further, by making the 
transformation $d\tau = dt/\sqrt{b}$ we can solve explicitly to obtain a general metric:
\be ds^2 = -(\tau^2-A) d\tau^2 + \f{\tau^2}{\tau^2-A} dx^2 + (\tau^2-A)^2(dy^2+dz^2) \ee
for which the energy density follows $(\tau^2-A)^{-2}$.
 
\subsection{Domain Walls}

The case of the domain wall \cite{Sikivie:1982qv,Lukas:1998yy} is also distinct from its isotropized counterpart. For a domain wall, the principal scale factor will describe the axis perpendicular to the wall
and the secondary scale factor those parallel to it \cite{Ipser:1983db}. In this case we see that $w_1=0$ and $w_2=-1$, again, choosing coordinates such that the domain wall is a non-
moving entity. Hence $F=7/2$, and we are in a region which is not described by the isotropic FLRW cosmologies - this solution falls into the class described by 
equation (\ref{anp}). The principal scale factor is then described by $a=a_o (A/\sqrt{b} + b^3) $ and hence close to the initial time singularity the dynamics is determined by
the value of $A$ - for positive $A$ there is a cigar like singularity, for zero $A$ a point-like, and for negative $A$ we see the principal scale factor vanish at a finite, non-zero
value of $b$ - a finite pancake. In this case, we find that our constant of motion is found following equation (\ref{alpha0}), and hence $b^{3/2}$ is proportional to time. Inverting this we find
the general metric for such solutions:
\be ds^2 = -dt^2 + \left(\f{A}{t^{\f{1}{3}}}+t^2\right)^2 dx^2 + t^{\f{4}{3}}(dy^2+dz^2) \ee
Therefore cosmologies dominated by domain walls always begin with a vanishing of the principal scale factor when
the secondary scale factor takes a finite value. 

The energy density of this situation is given by equation (\ref{Rhonp}), and we see that the value of $\rho$ at the 
initial singularity is not fixed by the presence of the domain wall itself, but actually depends on the free parameter determining the system, $A$. In the case where $A<0$ -
the finite pancake - the energy density is infinite at this surface, as it follows $\sqrt{b}$. When $A \geq 0$ (either as $A=0$, a point-like singularity, or $A>0$, a cigar) the energy
density near this initial singularity follows $\sqrt{b}$, and hence tends to zero. The general behaviour of the energy density is
\be \rho =\f{7 \sqrt{b}}{A+b^{\f{7}{2}}} = \f{7 t^{\f{1}{3}}}{A+t^{\f{7}{3}}} \ee

\subsection{Magnetic Fields}

Another case of interest is that of a stationary (primordial) magnetic field \cite{Barrow:1997mj}, which we shall choose to be aligned with the principal axis. In this case, the Maxwell tensor $F_i^j$ is zero
apart from $F_y^z=F_z^y=2B$, hence the stress-energy tensor is diagonal with entries $(B^2,-B^2,B^2,B^2)$ and so $w_1=-1$ and $w_2=1$ \cite{Jacobs:1969qca}. In the isotropic case, this
would be modelled through $w=1/3$ and hence have the same behaviour as radiation. However, the anisotropic nature of the field leads to a distinct behaviour. The 
behaviour of the principal scale factor is (up to a choice of constants) $a=\f{\sqrt{b-C}}{b}$ for a negative constant $C=-A$ of equation (\ref{anm}). Hence we can use our constant
of motion to show a general metric: Letting $d\tau=dt/b$ we can solve to find
\be ds^2 = -4(\tau^2+C)^2 d\tau^2 +\f{\tau^2}{(\tau^2+C)^2} dx^2 + (\tau^2+C)^2 (dy^2 + dz^2) \ee
and $t=\tau(\tau^2+3C)$ can be inverted to express in terms of $t$ as required. We see from this metric that the solution indeed has a barrel type singularity at $\tau \rightarrow 0$
at which $\rho \rightarrow C^{-1}$. This again constitutes an example of `matter that matters' as it has fundamentally changed the nature of the initial singularity. The general
form of the energy density is:
\be \rho= \f{C}{b^2} = \f{C}{(\tau^2+C)^2} \ee

\subsection{Relativistic Particles}

In contrast to the magnetic field case, there is another matter model for which the average pressure scales as one third of the energy density - a highly relativistic particle.
To model this situation, we consider a continuous, homogeneous stream of particles and align the principal axis with the direction of motion of the stream. In this case, we
see that the anisotropic pressures give $w_1=1$ and $w_2=0$, $F=2$ and so we are in the sector which is determined by equation (\ref{a1p}). We thus see the geometry
begin with an initial cigar like singularity (noting that $A$ must be positive due to the positivity of energy density) and the principal scale factor expands to infinity as the 
secondary scale factor expands.  In this case the constant of motion is $ab\dot{b}$, and so we can follow the same methods as above to find a solution. Setting 
\be d\tau = 2Ab^{\f{3}{2}} dt \ee
we can integrate to find that $b=\sqrt{\f{\log(\tau)}{A}}$ and $a=(\f{A}{\log(\tau)})^{\f{1}{4}} \tau$, and hence we have a general metric for solution which takes the form:
\be ds^2 = -\f{1}{4A^{\f{3}{2}} \sqrt{\log(\tau)}} d\tau^2 +\tau^2 \sqrt{\f{A}{\log(\tau)}} dx^2 + \f{\log(\tau)}{A}(dy^2+dz^2) \ee
which is valid for $\tau > 1$ and A positive. The energy density, given by equation (\ref{Rho1p}), is determined to be 
\be \rho = \f{4A}{b \exp(2Ab^2)} = \f{4A^{\f{3}{2}}}{\tau^2 \sqrt{\log(\tau)}} \ee
and hence is infinite at the initial singularity, and tends to zero as the space-time expands. 

\section{Discussion}\label{Discussion}

Various approaches to understanding the cosmological singularities of general relativity invoke the idea that the dynamical equations become ultra-local; spatial derivatives
contribute at a lower order than temporal derivatives \cite{Andersson:2004wp}, and thus the system is approximately described by a set of weakly interacting points each behaving as a homogeneous 
cosmology \cite{Garfinkle:2003bb,Berger:1998vxa,Berger:2004rz}. This has been used to justify the idea that quantum gravity itself may be a lower dimensional theory \cite{Carlip:2009km}, and has
been used to inform approaches to attacking generic singularities in quantum gravity \cite{Ashtekar:2008jb}. The strategy that is employed is to classify the nature of singularities (or their resolution) in
the set of homogeneous space-times that result from such a conjecture, and from there to argue that the fate of generic singularities is determined by the behaviour of these homogeneous models. Thus
a generic model of singularities in a quantum space-time can be obtained without having to deal with complexities that arise from spatial interactions. In particular the nature of the equations of motion in 
the homogeneous case is that the partial differential equations that determine solutions in general relativity become ordinary differential equations which are much more tractable for quantum considerations. 
Since the universe that results from these conjectures consists of a set of essentially independent points, each can be treated using methods of quantum mechanics rather than attempting a full quantum field
theory of gravity. 

Typically quantum models of cosmology it is assumed that the only relevant matter source is a stiff fluid, which can be modelled as a massless scalar field. This has provided the matter source in many of the quantum cosmological
treatments of Bianchi models \cite{Ashtekar:2009vc,WilsonEwing:2010rh}, in which it has been shown that that the classical singularity is overcome by the effects of quantum geometry, and it has been shown
that in fact such a treatment of homogeneous, isotropic cosmologies never gives rise to a singularity \cite{Singh:2009mz}. This result continues to hold in the anisotropic
sector \cite{Singh:2011gp}, but further investigation should be performed into the presence of anisotropic matter which dominates some singularities. 

Here we have studied the locally rotationally symmetric Bianchi $I$ and $VII_o$ models in which matter consist of a perfect fluid only, not considering homogeneous models in full generality.  Due to this
simplification the equations of motion became significantly more tractable than in the general case - the presence of curvature in other LRS Bianchi models would lead to a more complex 
replacement for the constant of motion $J$ which was used heavily in finding solutions. Further, when considering models which are not LRS the generic behaviour of one scale factor in terms of 
one of the others is not as clear cut - monotonicity is not guaranteed and identifying the full extent of solutions is not simple. The use of a single matter source can be viewed in a number of ways  - either
as an approximation describing the behaviour of a cosmology in which a single source dominates, or to be valid in the neighbourhood of a singularity in which once would expect that a single source (that 
with the greatest pressure in the contracting direction, say) would be dominant. In isotropic models, we see that the dominant energy density comes from those fields with the highest value of $w$ - those
whose pressures increase the most as space-time contracts to a point. However, in the presence of an anisotropic singularity, we should expect the dominant matter to be that which experiences not only
the greatest increase in energy density due to contracting directions, but also that which is diluted the least by expanding directions. It is therefore likely that anisotropic matter sources would play a more
significant role in such space-times. 

The matter that has been used as examples in this paper may not be that present at an initial singularity. Topological defects arise as a result of symmetry breaking in field theories as 
the temperature of the matter is reduced, and thus as density is increased we should expect symmetry to be restored. However we see that in the case of the domain wall there exist
solutions which both begin and end at singularities with zero energy density. Thus although there may be a global symmetry at a maximum of energy density (or above a certain value)
the behaviour near these singularities may indeed exist at broken states of the symmetry and thus exhibit topological defects. Further, although the matter can be considered to be a toy
model, there is as yet grand unified theory, and the nature of matter at high energy densities remains unknown. It is therefore logical to assume that there may be fields which are anisotropic
in their pressures, existing with preferred vectors (like the magnetic and electric fields) and it is conceivable that particles such as photons would remain all the way to the singularity, and thus
the approximation of isotropy in matter is unfounded. 

\section*{Acknowledgments}

The author is grateful to Rafael Alves Batista for useful discussions, encouragement to pursue a wide range of models and helpful comments. The author acknowledges helpful contributions from the referees of this paper whose input have added to the clarity of the work. This work was supported by a grant from the John Templeton Foundation. 

\bibliographystyle{ieeetr}
\bibliography{LRS}

\end{document}